\documentclass{article}
\usepackage{graphicx}
\usepackage{url}
\usepackage{latexsym}

\def\abstract#1{\vskip 7mm
        \begin{center}{\large Abstract}\par \smallskip
                \begin{minipage}[c]{12cm}
                        \small #1
                \end{minipage}
        \end{center}
}
\def\title#1{\begin{center}{\Large\bf #1}\end{center}}
\def\author#1{\vskip 5mm \begin{center}{#1}\end{center}}
\def\address#1{\begin{center}{\it #1}\end{center}}
\makeatletter
\@ifundefined{lesssim}{}{}
\@ifundefined{gtrsim}{}{}
\def\vereq#1#2{\lower3pt\vbox{\baselineskip1.5pt \lineskip1.5pt
\ialign{$\m@th#1\hfill##\hfil$\crcr#2\crcr\sim\crcr}}}
\makeatother

\begin{document}

\title{%
The Quintom Model of Dark Energy
 }
\author{%
Bo Feng$^{a,b,}$\footnote{E-mail:fengbo@resceu.s.u-tokyo.ac.jp}
 }
\address{%
$^{a}$ National Astronomical Observatories, Chinese Academy of
Sciences, Beijing 100012, P. R. China
\\ $^{b}$ Research Center for the
Early Universe(RESCEU), Graduate School of Science, The University
of Tokyo, Tokyo 113-0033, Japan }

\abstract{ In this paper I give a brief review on the recently
proposed new scenario of dark energy model dubbed $Quintom$. Quintom
describes the dynamical dark energy models where the equation of
state getting across the cosmological constant boundary during
evolutions. I discuss some aspects on the quintom model buildings
and the observational consequences. }

\section{Introduction}

    The nature of dark energy is among the biggest problems in modern
physics and has been studied widely. The simplest candidate of dark
energy is the cosmological constant, it suffers from the well-known
fine-tuning and coincidence problems~\cite{SW89,ZWS99}.
Alternatively, dynamical dark energy models with the rolling scalar
fields have also been proposed, such as
quintessence\cite{quint,pquint}, the ghost field of
phantom~\cite{phantom} and the model of k-essence which has
non-canonical kinetic term~\cite{kessence}.

Given that currently we know very little on the theoretical aspects
of dark energy, the cosmological observations play a crucial role in
our understandings. The model of phantom has been proposed in
history due to the fact that the observations have shown some mild
preference for an equation of state(EOS) smaller than
$-1$\cite{phantom}. Although in this scenario dark energy violates
the weak energy condition(WEC) and leads to the problem of quantum
instabilities\cite{Hawking:1985gh,Phtproblms}, we need more efforts
on this observation-inspired topic.

The Type Ia supernova (SNIa) observations from the HST/GOODS program
and the previous supernova data\cite{Riess04}, which make the only
direct measurements of dark energy, somewhat favor the dynamical
dark energy model with an equation of state getting across -1 during
the evolutions\cite{cooray}. If such a kind of dynamical dark energy
were verified by future observations, it would be a challenge to the
dark energy model buildings. Neither the cosmological constant nor
the dynamical scalar fields like quintessence or phantom would be
the source driving the current accelerated expansion of the
Universe. The model of quintessence has an equation of state which
is always no smaller than minus unity while the ghost field of
phantom has an EOS always no larger than -1. Basing on these facts
we proposed a new model of dark energy dubbed
$Quintom$\cite{Feng:2004ad}, in the sense that the required behavior
of the dynamical dark energy combines that of quintessence and
phantom. The purpose of this paper is to discuss briefly some
aspects on the quintom model buildings and the corresponding
observational consequences.

\section{The Quintom Model of Dark Energy}

Phenomenologically in comparison with quintessence and phantom, the
behavior of quintom is more flexible and it can lead to some
distinctive pictures in the determinations in the future of our
Universe. In Ref.\cite{Feng:2004ff} we proposed a scenario of
quintom with oscillating equation of state. We find oscillating
Quintom can unify the early inflation and current acceleration of
the universe, leading to oscillations of the Hubble constant and a
recurring universe. Our oscillating Quintom would not lead to a big
crunch nor big rip. In Ref.\cite{Xia:2004rw} we found interestingly
the current observations somewhat favor an oscillating quintom with
a much smaller period than that required in the recurrent universe
scenario. Note in Refs.\cite{Feng:2004ff,Xia:2004rw} they are
phenomenological studies only and are not focused on quintom model
buildings.

At the first sight it seems to be easy for the quintom model
buildings, since naively one may consider a model with a
non-canonical kinetic term with the following effective
Lagrangian\cite{Feng:2004ad}:
 \begin{equation}
\mathcal{L}=\frac{1}{2}f(T)\partial_{\mu}Q\partial^{\mu}Q-V(Q)~,
 \end{equation}
where $f(T)$  can be a dimensionless function of the temperature or
scalar fields. During the evolution of the universe when $f(T)$
crosses the point of zero it gives rise to the crossing of the
cosmological constant boundary. However one also needs to consider
the dynamics of the $f(T)$ term in reality and this makes it not
straightforward for successful quintom model buildings. When we
consider realistic quintom models we need also to consider their
spatial fluctuations. It is crucial for us to understand its
imprints in the cosmological observations. If we simply neglect dark
energy perturbations and start from parametrizations of the scale
factor $a(t)$, it would be very easy to construct quintom models
since this is somewhat like reconstruction of $a$ using $w(t)$.

It turns out that if we consider the usual kessence as the candidate
of quintom, at the crossing point it cannot be quantized in a
canonical way\cite{Zhao:2005vj}. Due to the problems on
perturbations, we cannot realize quintom with a single fluid or
single scalar field in the conventional way. In general one needs to
add extra degrees of freedom for successful quintom model buildings.
In Ref\cite{Feng:2004ad} we considered the simplest case with one
quintessence field and the other being the phantom field:
\begin{eqnarray}\label{double}
 \mathcal{L}& =& \frac{1}{2}\partial_{\mu}\phi_1\partial^{\mu}\phi_1-\frac{1}{2}\partial_{\mu}\phi_2\partial^{\mu}\phi_2
 \nonumber -V_0[{\rm exp}(-{\lambda\over m_p}\phi_1)+{\rm exp}(-{\lambda\over
 m_p}\phi_2)]~.
 \end{eqnarray}
Such a two-field model can easily cross the cosmological constant
boundary(See also \cite{Guo:2004fq}). However for the simplest
two-field model we are faced with the problem of ghost instabilities
inherited inevitably in the phantom
component\cite{Hawking:1985gh,Phtproblms}. Another possibility of
introducing the extra degrees of freedom for the realization of
quintom was proposed in Ref. \cite{Li:2005fm}, where we introduced
higher derivative operators to the Lagrangian. Specifically  we
considered a model with the Lagrangian
\begin{equation}\label{lagrangian} \mathcal{L}=-{1\over
2}\nabla_{\mu}\phi\nabla^{\mu}\phi+{c\over
2M^2}\Box\phi\Box\phi-V(\phi)~, \end{equation} where $\Box\equiv
\nabla_{\mu}\nabla^{\mu}$ is the d'Alembertian operator. The term
related to the d'Alembertian operator is absent in the quintessence,
phantom and the k-essence model, which is the key to make the model
possible for $w$ to cross over $-1$. We have shown in
\cite{Li:2005fm}
 this
Lagrangian is equivalent to an effective two-field
model\begin{equation}\label{alagrangian} \mathcal{L}= -{1\over
2}\nabla_{\mu}\psi\nabla^{\mu}\psi+{1\over
2}\nabla_{\mu}\chi\nabla^{\mu}\chi -V(\psi-\chi)-{M^2\over
2c}\chi^2~, \end{equation} with \begin{eqnarray}
\chi \equiv\frac{c}{M^2}\Box\phi\label{change}~,\\
\psi \equiv\phi+\chi~.  \end{eqnarray} Note that the redefined
fields $\psi$ and $\chi$ have opposite signs in their kinetic terms.
One might be able to derive the higher derivative terms in the
effective Lagrangian (2) from fundamental theories. For example it
has been shown that this type of operators appears as some quantum
corrections or due to the non-local physics in the string theory. In
principle as the Lagrangian (2) is equivalent to the two-field model
the phantom instabilities still exist. However we can also expect in
the two cases their behaviors are different when we consider the
possible interactions. In particular, Ref.\cite{Hawking:2001yt} has
shown that in this scenario ghosts arise because of the canonical
treatment, where $\phi$ and $\Box \phi $ are regarded as two
independent variables. An alternative quantization based on the path
integral seems to be intriguing towards solving the problem of
ghosts\cite{Hawking:2001yt}.

The model of quintom was initiated by the SNIa observations, in
realistic quintom model buildings we need to consider the imprints
in the concordance observational cosmology. In the probe of quintom
signatures in cosmic microwave background (CMB) and large scale
structure (LSS) we need to consider the effects of dark energy
perturbations. When dark energy is not simply the cosmological
constant in general it will cluster on the largest scales, which can
leave some imprints on the observations. Ref.\cite{Weller:2003hw}
has shown that for scalar models of dark energy like quintessence
and phantom, the effects of dark energy perturbations are to
introduce more degeneracies with the equation of state on CMB. In
our simplest two-field quintom case we have found that crossing the
cosmological constant boundary would not lead to distinctive
effects. On the other hand for models of scalar dynamical dark
energy the equation of state is not a constant in general, however
the effect on CMB can be almost identically described by a constant
EOS:
\begin{equation}\label{weff}
    w_{eff}\equiv\frac{\int da \Omega(a) w(a)}{\int da \Omega(a)}~~,
\end{equation}
hence it is easily understood when we include the effects of quintom
perturbations it will be in more degeneracy with the geometric
parameter $w$. In general when we add geometrical data the
degeneracy between a constant $w_{eff}$ and a dynamical $w$ can be
somewhat broken, but it still exists due to limits on the precisions
of the current SNIa observations.

In the study of quintom perturbations it is straightforward in the
two-field case. But cosmologists are sometimes more interested in
the inverse study on dark energy through fittings to the
cosmological observations. In the fittings one typically
parameterizes the equation of state. However when we parameterize
quintom-like dark energy the problem arises in the study on the
imprints of perturbations. We need to bear in mind what spectrum
this kind of dark energy displays: it cannot be a simple one-field
scalar or single ideal fluid, the perturbations will diverge for
these cases. We introduce a small positive constant $c$ to divide
the whole region of the allowed value of the EOS $w$ into three
parts\cite{Zhao:2005vj}: 1) $ w
> -1 + c$; 2) $-1 + c > w > -1 - c$; and 3) $w < -1 -c $. In Regions
1) and 3) the parameterized dark energy can be described as
conventional quintessence and phantom. In Region 2) numerically we
have set the derivatives of pressure and density perturbations to be
zero at the extremely limited matching point. Through this method
our study of parameterized quintom can resemble two-field quintom
models and no singularities appear. In Ref.\cite{Xia:2005ge} we have
made global fittings on the current status of dynamical dark energy
including quintessence, phantom and quintom. We have found that a
dynamical dark energy with the EOS getting across $-1$ is favored at
1$\sigma$ with the combined constraints from WMAP, SDSS and the
"gold" dataset of SNIa by Riess $et$ $al$, see Fig.1. In previous
investigations due to the problems on quintom perturbations the
fittings in the literature typically did not include dark energy
perturbations in the probe of dynamical dark energy. We can find
this will lead to nontrivial bias. Similarly we can also easily
understand quintom perturbation will play a significant role in
probing dynamical dark energy using future precise measurements like
SNAP and JDEM.

\begin{figure}[htbp]
\begin{center}
\includegraphics[scale=0.7]{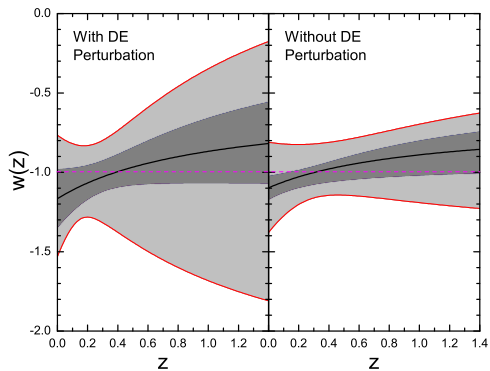}
\vskip-1.3cm \vspace{10mm}\caption{Constrains on w(z) using WMAP +
157 "gold" SNIa data + SDSS with/without DE
perturbation\cite{Xia:2005ge}. Median(central line), 68\%(inner,
dark grey) and 95\%(outer, light grey) intervals of w(z) using 2
parameter expansion of the EOS: $w(z)=w_{0}+w_{1}\frac{z}{1+z}$,
with $z$ being the redshift. \label{w}}
\end{center}
\end{figure}

We should point out currently on the observational aspect a
cosmological constant still fits the data quite well. Nevertheless
the SNIa observations, which make the only direct detections of dark
energy, somewhat favor a dynamical quintom-like dark energy. Note we
need to set several priors before we can get detailed investigations
on probing dark energy with SNIa only. On the other we need more
thorough understandings on the dynamical mechanism of Type Ia
supernova, both the quality and quantity of SNIa are to be improved.
Currently there are still various possibilities and alternatives on
the model of dark energy\cite{Li:2005zd}.

Both theoretical and observational probe of dark energy need still
go a long way. If we start always from a $\Lambda$CDM model in the
probe of our universe we cannot achieve more subtle physics beyond
that. This is necessary to bear in mind for us to understand the
nature of dark energy with the accumulation of the observational
data.

\section{Acknowledgments}
I thank my collaborators for discussions. I am indebted to Profs.
Xuelei Chen and Jun'ichi Yokoyama and without their kind hospitality
part of my work cannot be finished.


\begin{thebibliography}{99}


\bibitem{SW89}
S. Weinberg, Rev. Mod. Phys. {\bf 61}, 1 (1989). 

\bibitem{ZWS99}
I. Zlatev, L.-M. Wang, and P. J. Steinhardt, Phys. Rev. Lett. {\bf
82}, 896 (1999).


\bibitem{quint}
R.~D.~Peccei, J.~Sola and C.~Wetterich, Phys.\ Lett.\ B {\bf 195},
183 (1987); C. Wetterich, Nucl. Phys. B {\bf 302}, 668 (1988).

\bibitem{pquint}
B. Ratra and P. J. E. Peebles, Phys. Rev. D {\bf 37}, 3406 (1988).


\bibitem{phantom}
R. R. Caldwell, Phys. Lett. B {\bf 545}, 23 (2002).


\bibitem{kessence}
C. Armendariz-Picon, V. Mukhanov and P. J. Steinhardt, Phys. Rev.
Lett. {\bf 85}, 4438 (2000); T.~Chiba, T.~Okabe and M.~Yamaguchi,
Phys.\ Rev.\ D {\bf 62} (2000) 023511.


\bibitem{Hawking:1985gh}
  S.~W.~Hawking,
Print-86-0124 (CAMBRIDGE).


\bibitem{Phtproblms}
S.~M.~Carroll, M.~Hoffman and M.~Trodden, Phys. Rev. D {\bf 68},
023509 (2003); J. M. Cline, S.-Y. Jeon and G. D. Moore, Phys. Rev. D
{\bf 70}, 043543 (2004).


\bibitem{Riess04}
  A.~G.~Riess {\it et al.}  [Supernova Search Team Collaboration],
  Astrophys.\ J.\  {\bf 607}, 665 (2004).


\bibitem{cooray}
e.g. D. Huterer and A. Cooray, Phys. Rev. D {\bf 71}, 023506 (2005).


\bibitem{Feng:2004ad}
  B.~Feng, X.~L.~Wang and X.~M.~Zhang,
  Phys.\ Lett.\ B {\bf 607}, 35 (2005).

\bibitem{Feng:2004ff}
  B.~Feng, M.~Li, Y.~S.~Piao and X.~Zhang,
  arXiv:astro-ph/0407432.


\bibitem{Xia:2004rw}
  J.~Q.~Xia, B.~Feng and X.~M.~Zhang,
  Mod.\ Phys.\ Lett.\ A {\bf 20}, 2409 (2005).

\bibitem{Guo:2004fq}
  Z.~K.~Guo, Y.~S.~Piao, X.~M.~Zhang and Y.~Z.~Zhang,
  Phys.\ Lett.\ B {\bf 608}, 177 (2005); X.~F.~Zhang, H.~Li, Y.~S.~Piao and X.~M.~Zhang,
  arXiv:astro-ph/0501652.


\bibitem{Zhao:2005vj}
  G.~B.~Zhao, J.~Q.~Xia, M.~Li, B.~Feng and X.~Zhang,
  Phys.\ Rev.\ D {\bf 72}, 123515 (2005).


\bibitem{Li:2005fm}
  M.~Z.~Li, B.~Feng and X.~M.~Zhang,
  JCAP {\bf 0512}, 002 (2005).


\bibitem{Xia:2005ge}
  J.~Q.~Xia, G.~B.~Zhao, B.~Feng, H.~Li and X.~Zhang,
  arXiv:astro-ph/0511625.


\bibitem{Hawking:2001yt}
  S.~W.~Hawking and T.~Hertog,
  Phys.\ Rev.\ D {\bf 65}, 103515 (2002).

\bibitem{Weller:2003hw}
  J.~Weller and A.~M.~Lewis,
  Mon.\ Not.\ Roy.\ Astron.\ Soc.\  {\bf 346}, 987 (2003).

\bibitem{Li:2005zd}
e.g.  H.~Li, B.~Feng, J.~Q.~Xia and X.~Zhang,
  arXiv:astro-ph/0509272.



\end{thebibliography}
\end{document}